\documentstyle[prl,aps,amsfonts,twocolumn,epsfig,floats]{revtex}
\newcommand{\beq}{\begin{equation}}
\newcommand{\enq}{\end{equation}}
\begin{document}

\draft

\wideabs{

\title{Creation of a monopole in a spinor condensate}

\author{J.-P. Martikainen$^1$, A. Collin$^1$, and K.-A. Suominen$^{1,2}$}
\address{$^1$Helsinki Institute of Physics, PL 64, FIN-00014
Helsingin yliopisto, Finland\\
$^2$Department of Applied Physics, University of Turku, FIN-20014
Turun yliopisto, Finland}
\date{\today}

\maketitle

\begin{abstract}
We propose a method to create a monopole structure in a spin-1 spinor
condensate by applying the basic methods used to create vortices and
solitons experimentally in single-component condensates. We show, however,
that by using a two-component structure for a monopole, we can simplify our
proposed experimental approach and apply it also to ferromagnetic spinor
condensates. We also discuss the observation and dynamics of such a monopole
structure, and note that the dynamics of the two-component monopole differs
from the dynamics of the three-component monopole.
\end{abstract}
 }

\narrowtext

{\it Introduction}---The experimental realization of spinor Bose-Einstein
condensates~\cite{Stamper98,Barrett2001} makes it feasible to extend the
study of topological quantum objects into an entirely new field of physics.
Ordinary single-component condensates have many topologically interesting
properties such as the existence of vortices~\cite{Dalfovo1999}. But in
spinor condensates one can also study phenomena that cannot exist in the
single-component systems. One example is a monopole structure in a
spinor condensate with anti-ferromagnetic interactions as proposed by Stoof
{\it et al.} recently~\cite{Stoof2001}. Other novel possibilities also exist
such as skyrmions~\cite{Stoof2000,Khawaja2000}. With Bose-Einstein
condensates a crucial aspect is not only the existence and stability of
topological structures, but also the methods for their creation and
observation, as well as their dynamics. In this Letter we address all these
aspects for a monopole structure in an experimentally relevant case of a
multicomponent Bose-Einstein condensate.

A monopole is a topological defect in a vector field. It is characterized by
a unit vector that is radial in respect to some unique central point (i.e.
the ``hedgehog'' defect). In spinor condensates the vector quantity could be
the local spin of the condensed atoms~\cite{Garcia2000}, but other choices
are also possible~\cite{Busch99,Stoof2001}. Monopoles have been studied
theoretically in two-dimensional condensates~\cite{Garcia2000,Busch99}.
Recently some results for the three-dimensional case (such as density
distribution, energy and dynamics of such a defect) have been studied by
Stoof {\it et al.}~\cite{Stoof2001}.  They described a monopole created in
an antiferromagetic spin-1 condensate such as $^{23}{\rm Na}$. The monopole
was characterized by a spinor
\begin{eqnarray}
     \label{spinor}
     \zeta =\sqrt{\frac{n}{2}}\left( \begin{array}{c}
     -m_{x}+im_{y}\\
     \sqrt{2}m_{z}\\
     m_{x}+im_{y}
     \end{array}\right),
\end{eqnarray}
where $n$ is the condensate density and the vector ${\bf m}=\pm{\bf r}/r$ is
a radial unit vector  and has the spherically symmetric ``hedgehog''
structure.  Stoof {\it et al.} demonstrated that this particular spin
texture is a unique consequence of the unit winding number and minimization
of the gradient energy. The monopole can also be displaced from the center
of the trap without changing their argument. The spinor in Eq.~(\ref{spinor})
is non-magnetized and can be achieved from the single-component mean-field
ground state, $\zeta_0^T=\sqrt{n}\left(0\, 1\, 0\right)$,
with local spin-rotations.
Consequently, at each position it resembles the ground state and thus the
absence of dynamical instabilities which
lead to domain formation~\cite{Miesner1999,Stenger1999} is ensured.

{\it Two-component monopole}---One is not, however, limited to the
antiferromagnetic texture given in Eq.~(\ref{spinor}) when considering
monopoles. We can alternatively map the vector ${\bf m}$ into an
effective two-component system:
\begin{eqnarray}
     \label{two_comp}
     \zeta =\sqrt{n}\left( \begin{array}{c}
     -m_{x}+im_{y}\\m_{z}\\0
     \end{array}\right).
\end{eqnarray}
To ensure the stability of this texture against
phase-separation~\cite{Timmermans98}, the spin-1 condensate must have
ferromagnetic interactions, which makes the $^{87}{\rm Rb}$ spinor
condensate a potential candidate. In other words, the preparation of a
monopole is not limited to the antiferromagentic $^{23}{\rm Na}$ system as
expected before~\cite{Stoof2001}, if we accept spinors that do not have
the order-parameter space of the ground state.
Consequently, it should be noted, that the
texture in Eq.~(\ref{two_comp}) can not be produced by local rotations from
the ferromagnetic ground state, $\zeta_F^T=\sqrt{n}\left(1\, 0\, 0\right)$.

The static properties of the two-component monopole are similar to those
of a three-component spinor (there are differences in dynamics, as we shall
discuss later). Because it should also be easier to create experimentally, we
focus on the two-component case although our calculations are done for the
actual three-component system. We note that as the spinor in
Eq.~(\ref{two_comp}) is not the mean-field ground state locally, some
relaxation towards the true ground state is to be expected. But this
requires spin-changing processes, that are very slow and can thus be
ignored at timescales of interest.

We describe the spinor condensate with a multicomponent wave function and
label the components as $\psi_m$ with the spin projection quantum number $m$
($m=0,\pm 1$). The mean-field Gross-Pitaevskii (GP) equations are~\cite{Pu99}
\begin{eqnarray}
     i\hbar\frac{\partial\psi_{-1}}{\partial t}&=&{\cal L}\psi_{-1}+
     \lambda_a\left(\psi_0^2\psi_1^*+|\psi_{-1}|^2\psi_{-1}+
     |\psi_0|^2\psi_{-1}\right.\nonumber\\
     &&\left.-|\psi_1|^2\psi_{-1}\right)\nonumber\\
     \label{GPs}
           i\hbar\frac{\partial\psi_{0}}{\partial t}&=&{\cal L}\psi_{0}+
     \lambda_a\left(2\psi_1^*\psi_{-1}^*\psi_0+|\psi_{-1}|^2\psi_{0}+
     |\psi_1|^2\psi_{0}\right)\\
         i\hbar\frac{\partial\psi_{1}}{\partial t}&=&{\cal L}\psi_{1}+
     \lambda_a\left(\psi_0^2\psi_{-1}^*+|\psi_1|^2\psi_{1}+
     |\psi_0|^2\psi_{1}\right.\nonumber\\
     &&\left.-|\psi_{-1}|^2\psi_{1}\right)\nonumber,
\end{eqnarray}
where ${\cal L}=-\frac{\hbar^2}{2m}
\nabla^2+V_{trap}+\lambda_s\left( |\psi_{-1}|^2+|\psi_{0}|^2+
|\psi_{1}|^2\right)$, $\lambda_s=\frac{4\pi\hbar^2}{m} (a_0+2a_2)$,
$\lambda_a=\frac{4\pi\hbar^2}{m}(a_2-a_0)$, and $a_F$ is the s-wave
scattering length in the total hyperfine two-atom $F$-channel. For
a cylindrically symmetric trapping potential we use $V_{trap}=
m\omega_r^2(x^2+y^2)/2+ m\omega_z^2 z^2/2$.

\begin{figure}[bht]
\centerline{\epsfig{file=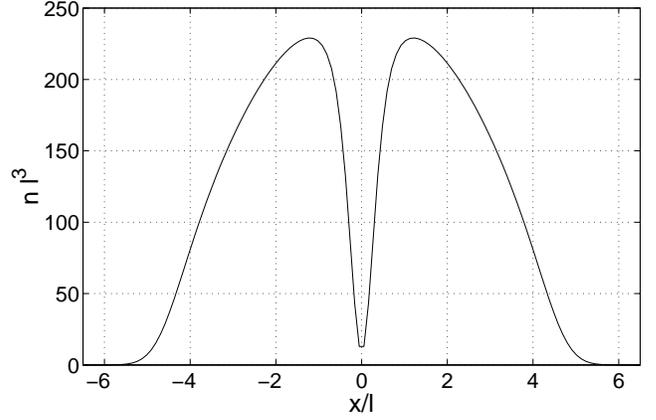,width=8.5cm}}
\vspace*{0.5cm}
\caption[fig1]{
The total density ($l=\sqrt{\hbar/m\omega}$) of the spin-$1$ condensate with
$5\cdot 10^4$ $^{87}{\rm Rb}$ atoms when $y=z=0.05$ (origin is not present
in our discretization scheme). The monopole is at the center of a
spherically symmetric trap with a trap frequency
$\omega=\omega_r=\omega_z=(2\pi)\,50$ Hz.
\label{ground_density}}
\end{figure}

{\it Monopole structure and stability.}---To understand and test our approach
for monopole creation we study the monopole structure by solving
Eq.~(\ref{GPs}) numerically. In order to create the monopole as an initial
state of our numerical study we force the texture given in
Eq.~(\ref{two_comp}) (or Eq.~(\ref{spinor})) into the order parameter
(this should not be confused with the actual proposal for experimental
creation of the monopole, to come later). The imprinted spinor is then
propagated in imaginary time until sufficient convergence is reached.
If the monopole is at the center of the condensate the imprint has to be
done only once at the beginning of the iteration, otherwise the imprint must
be repeated in the course of the iteration to prevent  the monopole from
drifting away from the intended location, to a location with lower energy.
In Fig.~\ref{ground_density} we show the typical density distribution of
the spin-1 monopole located at the center of a trap.

By looking at the individual condensate components we gain relevant insight
into the structure of the monopole. Here $\psi_{-1}$ has a vortex at $z=0$ with
a core size that is a function of $z$. On the other hand, $\psi_0$ goes
through a $\pi$ phase shift as we move from positive to negative $z$ values;
consequently, this component relates to a soliton in a single-component
condensate. The $m=0$ component atoms fill the vortex line everywhere else
except at the origin, where the density of $m=0$ component also vanishes.
Therefore the intersection of the vortex line with the soliton plane gives
rise to a monopole core. Thus, if we can experimentally create a vortex and
a soliton in a two-component system, we can obtain a monopole.

If we displace the monopole from the center of the spherically symmetric
trap, the energy of system decreases as function of the displacement
(in qualitative agreement with the results in  Ref.~\cite{Stoof2001}). This
indicates that in a dissipative environment the monopole is expected to
drift to the edge of the condensate and vanish.  The estimate for the
timescale of this process goes beyond the model used in this
paper. But in single-component condensates the vortex lifetimes can be
several seconds~\cite{Madison2000} and this timescale is also, presumably,
indicative of the monopole lifetime. On the other hand, the dissipative
processes might be masked by the topological instability of the monopole.
The unit winding number of the spin texture is not sufficient to protect the
topological stability of the system and instead of drifting smoothly to the
condensate edge, the monopole might decay into other excitations.

\begin{figure}[bht]
\centerline{\epsfig{file=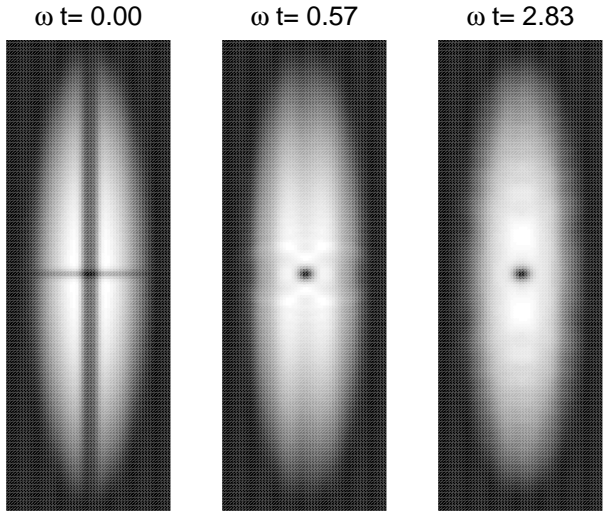,width=8.0cm}}
\vspace*{1cm}
\caption[fig2]{
The time evolution of the total density of a condensate with $5\cdot 10^4$
$^{87}{\rm Rb}$ atoms in a cylindrically symmetric trap with frequencies
$\omega=\omega_r=(2\pi)\,250$ Hz and $\omega_z=(2\pi)\,50$ Hz.
Initially there is a vortex at the $m=-1$ component (perpendicular line)
and a soliton at the $m=0$ component (horizontal line). The vortex and 
the soliton are imprinted at $t=0$
and the figures show the cut of the total density in $y=0.09$ plane.
$x$-axis is along the horizontal direction.
\label{Monopole_Creation}}
\end{figure}

{\it Creation of monopoles.}---The separate look into each spin
component of the monopole structure suggests a possible way to create it.
For example, we can prepare a spinor condensate with $2/3$ population at the
$m=1$ state and the rest at $m=0$ state, e.g. with an
rf-pulse~\cite{Mewes97}.  A ``blueprint'' of the monopole is achieved by
creating a vortex into the $m=1$ component and a soliton (with $\pi$ phase
discontinuity) into the $m=0$ component, both at the trap center. The vortex
line should lie along the phase discontinuity in the $m=0$ wavefunction. In
Fig.~\ref{Monopole_Creation} we demonstrate the time-evolution of such a
mixture in a cigar shaped trap in real time, when the soliton and vortex
were created using phase-imprint method~\cite{Dobrek1999}. With other
excitations abound, it is clear that we nevertheless have a monopole inside
the condensate.

In our numerical studies we have used a certain amount of smoothing to
reduce the amount of noise that would be created if a phase-imprint is too
abrupt. Smoothing for a vortex was done by assuming that not only do we
have a phase-mask, but also a narrow (on the order of the coherence length)
laser beam that bores a hole through the $m=1$ component along the vortex
line. For a soliton the $\pi$ phase jump was done at the distance of the
order of the coherence length. Without such smoothing  our numerical
approach becomes unstable. It is not clear how much is  this smoothing
required in actual experiments,  although some amount would be always
present. To obtain this combined phase-imprint  experimentally is
probably complicated, but at least the two main ingredients, namely
experimental creation of vortices~\cite{Matthews1999} and
solitons~\cite{Burger1999} has already been achieved.

{\it Observation of monopoles.}---The monopole core has roughly the size of
the healing length, and it is inside the condensate. Thus its direct
observation is difficult.  But the same methods used to observe
vortices~\cite{Madison2000,Matthews1999} and vortex
rings~\cite{Anderson2000} can be applied to observe monopoles as well. One
should first let the condensate expand and then image the 3D structure of
the different $m$-states using two orthogonal probe beams. As for separating
the different $m$-states, one can use an appropriate Stern-Gerlach
apparatus~\cite{Miesner1999}.

We have studied the behavior of a freely expanding monopole using the
time-dependent generalization of the monopole GP-equation in
Ref.~\cite{Stoof2001}. This approximation assumes equal scattering lengths,
but at the timescales of interest the role of differing scattering lengths
is in fact negligible. In the limit that $a_2=a_0$ all the components feel
the same spherically symmetric potential (external potential and the mean
field terms) and are not sensitive to the phase of the other components.
Therefore a single-component GP-equation is sufficient for modelling the
expanding monopole.

\begin{figure}[bht]
\centerline{\epsfig{file=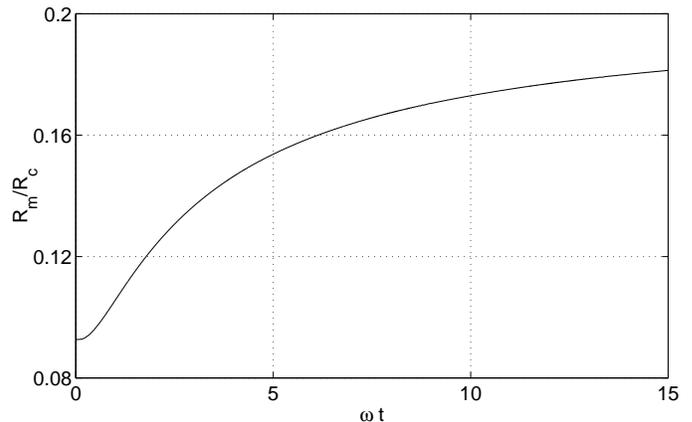,width=9.0cm}}
\caption[fig3]{
Time evolution of the ratio of the monopole core size $R_m$ to the
condensate size $R_c$. These sizes were determined from locations where the
density was one half of the maximum density. Initially we assumed
$5\cdot 10^4$ condensed rubidium atoms in a spherically symmetric trap with
a trap frequency $\omega=(2\pi)\,50$ Hz.
\label{Expansion_fig}}
\end{figure}

In Fig.~\ref{Expansion_fig} we show an example of the time-evolution of the
ratio of the  monopole core size to the system radius once the trapping
potential is turned off. Expansion is qualitatively similar to vortex
expansion in a scalar condensate. At small times the monopole core size,
$\xi$, adjusts  (almost) instantaneously to the local
density~\cite{Lundh1998} and one expects the size of the core to scale as
the healing length, $\xi_0=1/\sqrt{4\pi an}$. If we model the wave function as
\beq
	  \Psi(r)=A\exp\left(-\frac{r^2}{2R^2}\right)
	  \exp\left(\frac{i\beta r^2}{2}\right),
\enq
where $A$ is the normalization factor, the monopole grows faster than the
expanding  condensate, or more precisely
\beq
	  \frac{\xi_0}{R}=\frac{\pi^{1/4}R^{1/2}}{\sqrt{4\pi a N}}.
\enq
This happens as long as the characteristic time for adjustment of the core
size, $\tau_{ad}\sim \hbar/n\lambda_s$, is much less than the expansion
time $\tau_{ex}\sim R/c_s$, where $c_s$ is the sound velocity. The parameters
in Fig.~\ref{Expansion_fig} imply that in this  regime the condensate can
expand by an order of magnitude, and in the end of this regime the
monopole size compared to the condensate size has greatly increased. At
later times the atoms in the cloud will evolve as free particles and
$\xi_0/R$ will settle to some constant value. We also compared the results
obtained with a single component GP-equation (at small times) against the
solution of the multi-component GP-equations~(\ref{GPs}) and found that the
two approaches give essentially the same results.

{\it Monopole dynamics}---The dynamics of a monopole is quite interesting.
As expected, the monopole at the origin is stable and stationary. A displaced
monopole, on the other hand, behaves differently~\cite{own_comment}. The
monopole precesses around the trap center inside the condensate, just like a
displaced vortex line does. It returns to its initial location after
$T=2\pi/\Omega_P$. For a vortex close to the center the precession
frequency $\Omega_P$ coincides with the frequency of the anomalous mode
$\omega_a$. For a disc-shape trap an analytic result is
available~\cite{Svidzinsky2000} and is given by
\beq
     \label{anomalous}
	  \omega_a=-\frac{3}{2}\frac{\hbar}{mR^2}\ln
\left(\frac{R}{\xi}\right),
\enq
where $R$ is the radius of the system and $\xi$ is the healing length.
Even though the trap geometry in our example is nowhere near the disk-shape
we expect that Eq.~(\ref{anomalous}) gives a reasonable order of magnitude
estimate. Especially so since the $m=-1$ atoms are "squeezed" between lobes
of $m=0$ atoms, thus making the disc-shaped approximation rather justified.

As a test case we take a spherically symmetric trap with trap frequency
$\omega=(2\pi)\, 50$ Hz and a $^{87}{\rm Rb}$ condensate with
$5\cdot 10^4$ atoms. Setting $R$ to the Thomas-Fermi radius of the system
and calculating $\xi$ from the Thomas-Fermi result at the trap center we
get an estimate $\omega T\approx 35$. This value is fairly close to the
value actually seen in our 3D simulation of the monopole dynamics.
If the monopole was displaced by one-fifth of the Thomas-Fermi radius
our numerical result is $\omega T\approx 38$. The above estimate is
surprisingly accurate. In particular because the precessing vortex should
feel the mean field of the other component.

The dynamical behavior of the three-component monopole of Eq.~(\ref{spinor})
is different from above. In this case one has a vortex at the $m=-1$ state
and  an anti-vortex at the $m=1$ state. As a displaced vortex and an
anti-vortex precess in opposite directions, the monopole core will vanish
only to reappear at the opposite side as soon as the vortices have precessed
that far. This recurrence is almost perfect~\cite{own_comment}. Partial
revival of a monopole has also been predicted in case of a 2D
monopole~\cite{Busch99}. As the vortices precess in opposite directions the
order parameter becomes magnetized and can no longer be represented in the
form given by Eq.~(\ref{spinor}). Therefore, the order parameter is no
longer in the order parameter space of the ground state. Obviously, our
proposed approach to create monopoles experimentally applies to the
three-component case as well. But then one needs to create a
vortex/antivortex in the $m=\pm 1$ state, respectively, in addition to the
soliton in the $m=0$ state.

In Ref.~\cite{Stoof2001} the dynamics of the monopole were due to the  two
different scattering lengths. In our inhomogeneous spinor condensate the
dynamics are not intimately connected with differing scattering lengths.
Setting $a_2=a_0$ does not change our results qualitatively and even the
quantitative changes are small. At longer times some population dynamics can
be observed, but the dynamics of the total density is almost unaffected.
Therefore it seems clear that in an inhomogeneous spinor-condensate the
dynamical behavior of the monopole goes beyond the model suggested by Stoof
{\it et al.} in Ref.~\cite{Stoof2001}.

To summarise, we have proposed and demonstrated numerically a method to
create monopoles in three-dimensional Bose-Einstein condensates, and shown
that monopole creation is not limited to the antiferromagnetic spinor
condensates. In addition we have studied the detection by expansion of such
monopoles, and also the dynamics of displaced monopoles in a trap. For the
creation of two-dimensional monopoles, an off-resonant Raman beam has been
suggested~\cite{Garcia2000}, but so far there has not been any suggestions
for creating a three-dimensional monopole in a realistic experiment. Also,
by replacing a single vortex at one component with a lattice of vortices in
our approach could lead to creation of multiple monopoles, and allow one to
study interactions between monopoles.

Authors acknowledge the Academy of Finland (project 43336) and the National
Graduate School on Modern Optics and Photonics for financial support.
Discussions with J. Calsamiglia, U. Al Khawaja, M. Mackie and J. Piilo are
greatly appreciated.

\end{document}